\documentclass[conference]{IEEEtran}
\IEEEoverridecommandlockouts
\usepackage[caption=false,font=footnotesize]{subfig}
\usepackage{cite}
\usepackage{amsmath,amssymb,amsfonts}
\usepackage{graphicx}
\usepackage{textcomp}
\usepackage{xcolor}
\def\BibTeX{{\rm B\kern-.05em{\sc i\kern-.025em b}\kern-.08em
    T\kern-.1667em\lower.7ex\hbox{E}\kern-.125emX}}

\usepackage{algorithm}
\usepackage{algpseudocode}
\algrenewcommand\algorithmicrequire{\textbf{Input:}}
\algrenewtext{Loop}{\textbf{loop:}}
\algrenewtext{If}[1]{\textbf{if} #1\textbf{:}}
\algrenewtext{Else}{\textbf{else:}}
\algrenewtext{Function}[2]{\textproc{#1}(#2)\textbf{:}}
\algtext*{EndLoop}
\algtext*{EndIf}
\algtext*{EndFunction}

\begin{document}

\title{Tags Helping Tags: Coordination-Free Power Boosting for Battery-less Backscatter Networks}

\author{\IEEEauthorblockN{Mathew Salas\IEEEauthorrefmark{1}, Xin Xiong\IEEEauthorrefmark{1}, Sihua Shao\IEEEauthorrefmark{1}, Atef Elsherbeni\IEEEauthorrefmark{1}, Mostafa Hassanalian\IEEEauthorrefmark{2} and Rafael Fierro\IEEEauthorrefmark{3}}
\IEEEauthorblockA{\IEEEauthorrefmark{1}Department of Electrical Engineering,
Colorado School of Mines, Golden, CO 80401 USA
\\\IEEEauthorrefmark{2}Department of Mechanical Engineering, New Mexico Tech, Socorro, NM 87801 USA
\\\IEEEauthorrefmark{3}Department of Electrical and Computer Engineering, University of New Mexico, Albuquerque, NM 87131 USA \\Emails: mathew\_salas@mines.edu, xin\_xiong@mines.edu, sihua.shao@mines.edu, aelsherb@mines.edu, \\mostafa.hassanalian@nmt.edu, rfierro@unm.edu}
\vspace{-30pt}
\thanks{This work was supported by the National Science Foundation under Award Nos. CNS-2430772 and CNS-2431272. (\textit{Corresponding author: Sihua Shao}).}
}

\maketitle

\begin{abstract}
Battery-less backscatter sensor nodes powered by radio-frequency wireless power transfer are attractive for large or hard-to-reach deployments, but their range is typically limited by the forward power-transfer link rather than by communication sensitivity. We present a method for improving power delivery, and thus range, in dense tag deployments by having nearby tags act as helpers: each helper terminates its antenna in purely reactive impedances, reflecting the power beacon's signal toward a target tag so that it combines constructively with the direct line-of-sight path and increases the power the target harvests. On a hardware testbed, helpers using discrete impedance control are shown to increase the target tag's harvested power by over 50\% above an all-absorbing baseline, while randomly set helper impedances give no reliable benefit, and are sometimes even detrimental. We further propose and implement a local algorithm that lets each helper tune its own impedance using only measurements of the target's packet rate, without channel estimation or coordination between helpers, and we show that it brings the target’s packet rate close to that of an independently tuned configuration. These results suggest that independent, passive sensor nodes can cooperatively extend each other's power-transfer range with minimal added infrastructure.
\end{abstract}


\section{Introduction}
Battery-less wireless sensor nodes are attractive for deployments that are large or hard to reach, where replacing or recharging batteries is impractical and discarded batteries carry an environmental cost. Recent work has even envisioned dropping large numbers of such devices from a drone and letting the wind disperse them \cite{iyer2022}. These nodes must harvest their energy from somewhere, and most options depend on a particular energy source being present at usable levels. Solar harvesting degrades at night, under cloud cover, or when a panel is obscured, and thermal, vibration and ambient-RF harvesting each assume the environment happens to supply that energy. Light-based wireless power transfer has also been explored \cite{salustri2024}, but shares many of the issues of solar harvesting and requires a clear line of sight. Radio-frequency wireless power transfer (RF WPT) removes this assumption by using a dedicated continuous-wave (CW) power beacon. Because backscatter communication, suitable for the low-power requirements of battery-less devices, already requires an RF carrier, the energy source is present in essentially every scenario where backscatter works. This makes RF WPT a natural fit for battery-less backscatter sensor networks.

The difficulty is range. In passive backscatter systems, the forward power-transfer link typically restricts range more than the receiver's sensitivity does \cite{boaventura2025}. The effect is gradual in a tag whose operating period is tied to its harvested power. As the tag moves farther from the beacon, its harvested power drops and so does its packet rate. Eventually the harvested power falls below the tag's leakage, quiescent and startup consumption, and the tag can no longer operate. The obvious remedies are unattractive. Raising the beacon's transmit power is limited by transmit-power constraints and the cost of high-power RF hardware. Adding beacons or moving them closer is inefficient because a beacon is generally far more expensive than a sensing node. High-gain or beamforming antennas narrow the beam, so each beacon covers a smaller region.

Dense deployments offer a different opportunity. Swarm-style sensing and applications that need high spatial diversity already place many tags near one another. Rather than treating neighboring tags as clutter, we ask whether they can help each other. A backscatter tag terminated in a purely reactive impedance reflects the incident wave with a phase determined by that impedance. A tag acting as a \textit{helper} therefore creates a second path from the beacon to a \textit{target} tag, in addition to the direct line-of-sight path. If the two signals arrive in phase, they add constructively, the signal at the target strengthens, and its harvested power increases. If they arrive out of phase, they subtract. The helper's impedance choice decides which happens. Serving as a helper has a cost, since a reflecting tag cannot harvest, sense or communicate for itself. The trade-off can still be worthwhile. It can bring an otherwise out-of-range target into operation. It can also shift service toward a target whose data matters, for example for fairness, by having nearby tags that already have surplus energy spend time helping rather than transmitting.

This idea is related to, but distinct from, several lines of work. Collaborative backscatter recruits helper tags to strengthen tag-to-tag communication links \cite{ahmad2021,ahmad2025}, with reflection phases set from measured channel phases \cite{ahmad2021}. We instead target the power available at the target tag on the beacon-to-target link. Reconfigurable intelligent surface (RIS)-aided WPT also shapes reflections to raise harvested power \cite{wu2019,Ramezani2021}, but it relies on dedicated multi-element surfaces optimized jointly. Our helpers are identical, independent sensor nodes with their own sensing function, and each has only a few discrete impedance states. Conceptually, the closest analogue is distributed beamforming, in which independent nodes adjust their phases using feedback about the received signal \cite{mudumbai2010}. Here the reflecting nodes are passive, and no feedback message is needed, because each helper can infer the target's harvested power from the target's own packet rate.

Setting the helpers' impedances by hand does not scale, and the best setting depends on geometry and on an environment that may change slowly. We therefore also develop a way for helpers to find good impedance values by themselves, using only local measurements and processing.

The contributions of this paper are as follows:
\begin{itemize}
    \item Feasibility of helper-assisted power transfer. On a hardware testbed, we show that helper tags using discrete impedance control can induce constructive interference at a target tag's harvester. With three helpers, harvested power increased by 55.8\% over an all-absorbing baseline (from 37 to 58 $\mu$W) in one geometry.
    \item Local, autonomous impedance selection. We propose and implement a novel algorithm that each helper runs on its own microcontroller (MCU). It uses only the target's observed packet rate, measured through the helper's own harvester/receiver circuitry, to raise the target’s packet rate toward that of an independently tuned configuration, with the median trial going from about 50\% to about 95\% of the tuned optimum, a 1.9× rate increase. Eight of 12 trials end at 90\% or more.
\end{itemize}

\section{SYSTEM MODEL AND TESTBED}
\subsection{Overview and Notation}
\begin{figure}[ht]
    \centering
    \subfloat[\small Testbed]{%
        \includegraphics[width=0.53\linewidth]{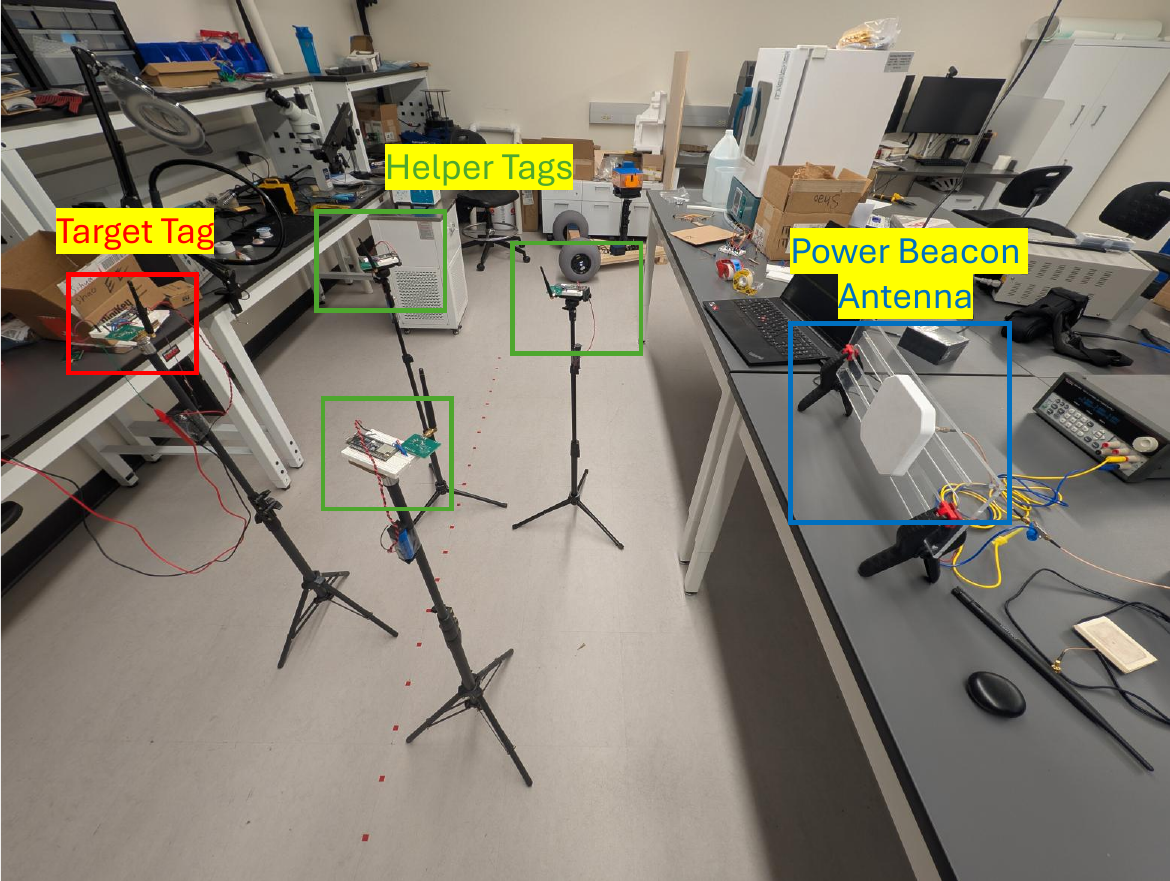}}
    \hfill
    \subfloat[\small RF Frontend]{%
        \includegraphics[width = 0.46\linewidth]{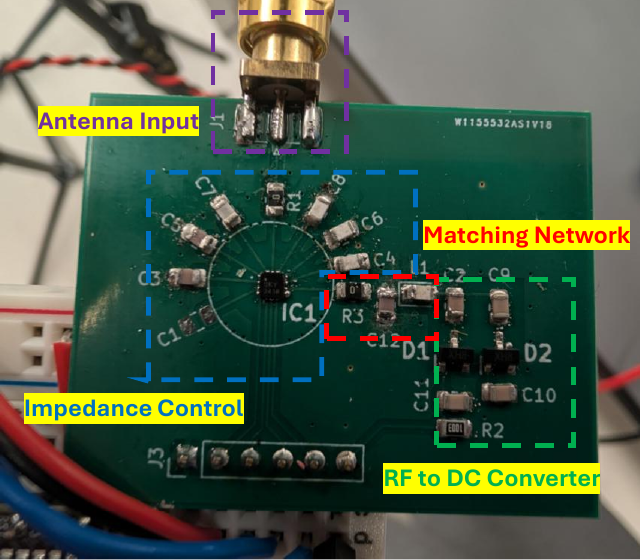}}
    \caption{(Left) Photo of testbed showing target tag (red), helper tags (green), and power beacon (blue). (Right) Photo of custom RF frontend PCB showing the antenna input (purple), the impedance switch (blue), the matching network (red), and the 2-stage RF-to-DC converter (green).}
    \label{fig:testbed}
\end{figure}

We consider a CW power beacon, a target tag whose harvested power we want to increase, and $N$ helper tags placed near the target. All tags share the same RF frontend. Each tag connects its antenna to one of eight terminations. State $8$ is the absorbing state, in which the antenna feeds the matching network and RF-to-DC converter. States $k=1,\dots,7$ are reflecting states (open circuit or one of six reactive loads), with reflection coefficient $\Gamma_k=|\Gamma_k|e^{j\phi_k}$. Due to real-world non-idealities, such as imperfect matching or internal resistance, the actual magnitude of the reflection coefficient is around $0.8$ to $0.9$; however, we approximate it as $|\Gamma_k|\approx 1$. We model a helper in state $8$ as not contributing to the field at the target. A scenario with fewer than $N$ active helpers is therefore obtained by placing some helpers in state $8$. We assume the propagation environment and tag positions remain stable over the time needed to tune a helper. Fig. \ref{fig:testbed} shows the experimental testbed, including custom PCBs for the RF frontend. The MCU and other hardware are set up on breadboards, and each tag is mounted on a tripod.

\subsection{Tag Hardware}
\begin{figure}
    \centering
    \includegraphics[width=0.8\linewidth]{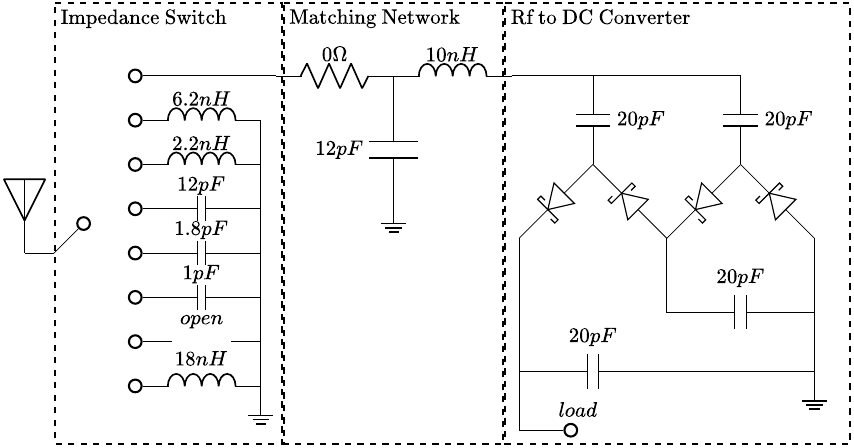}
    \caption{Circuit diagram of the RF frontend, including antenna, impedance switch, matching network, and 2-stage Dickson RF-to-DC converter.}\vspace{-10pt}
    \label{fig:RFFrontend}
\end{figure}

Each tag (Fig. \ref{fig:RFFrontend}) connects a 2.15 dBi microstrip dipole antenna by SMA to an RF frontend board (component details in Table~\ref{tab:testbed}). The board contains a single-pole eight-throw (SP8T) switch that selects the termination, a matching network, and a two-stage Dickson RF-to-DC converter. The seven reflecting terminations are an open circuit and six capacitive or inductive loads. Their initial values were computed by solving \eqref{eq:reflection} for the reflection coefficient of each candidate impedance
\begin{equation}
    \label{eq:reflection}
    \Gamma = \frac{Z - Z_{in}}{Z + Z_{in}},
\end{equation}

where $Z_{in}$ is the input/antenna impedance, $Z$ is the target/load impedance including the reactive loads and the RF-to-DC converter impedance. $\Gamma$ is the reflection coefficient.

The values were then tuned heuristically against vector network analyzer (VNA) measurements to spread the reflection phase around the circle. Table~\ref{tab:reflections} lists the final states. Adjacent spacings range from 33° to 81° (mean $\sim51$°), so even the worst-case phase error is within about $81^\circ/2 \approx 41^\circ$ of an available state. These values were measured against a 50 $\Omega$ VNA reference. With the real antenna, which is not perfectly matched, they will shift slightly.

An on-board MCU controls the switch. In the feasibility experiments (Section IV-A), all tags are battery powered (ESP32-S3), and only the target's converter output is observed. For the algorithm evaluation (Section \ref{sec:algorithm experiment}), the target is a fully power-cycling tag (Fig. 3). The Dickson output charges a storage capacitor, and an low-dropout (LDO) regulator based hysteresis circuit \cite{salas2025} powers an ATtiny20 MCU when the capacitor reaches a turn-on voltage and cuts it off at a lower turn-off voltage. Each time the MCU is powered, it transmits a short packet (preamble, sync word, one data byte) by toggling the switch to generate an on-off keying (OOK) signal, and the cycle repeats. The interval between packets therefore tracks the harvested power. Helpers remain battery powered. In the absorbing state, a helper samples its converter output with the MCU's analog-to-digital converter (ADC) and times the target's packets; see Section III for details.

\begin{figure}
    \centering
    \includegraphics[width=\linewidth]{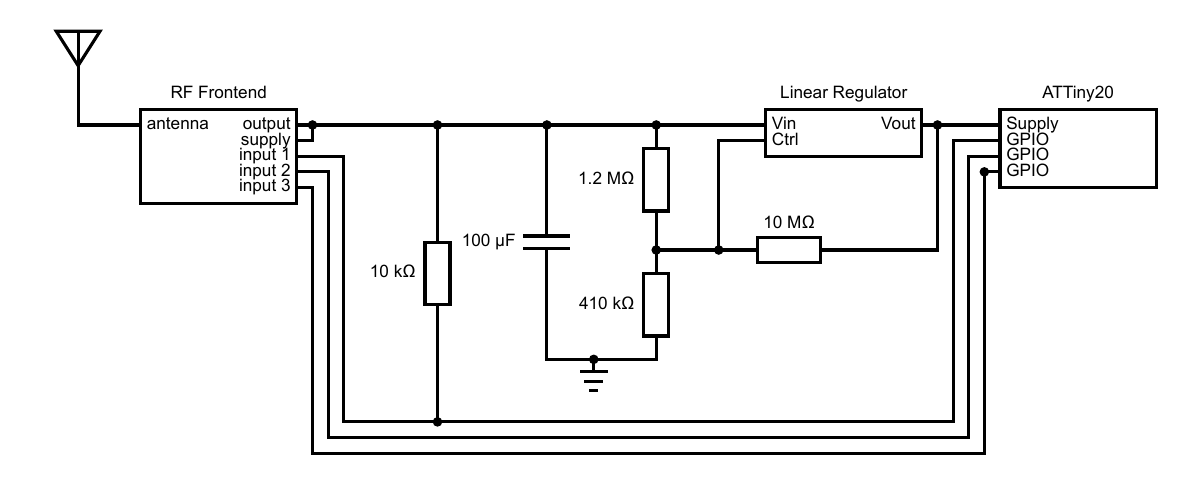}\vspace{-10pt}
    \caption{Diagram of the battery-less tag design, used for algorithm testing.}\vspace{-10pt}
    \label{fig:batteryless}
\end{figure}

\subsection{Beacon, Geometry, and Measurement}

The beacon is a USRP N210 controlled by a minimal GNU Radio flowgraph that produces an unmodulated 917 MHz tone. The tone is amplified and radiated at 22 dBm from either a 5.5 dBic circularly polarized patch antenna for the feasibility experiments, or an 8.5 dBic circularly polarized patch antenna for the algorithm evaluation (Table~\ref{tab:testbed}). The target is fixed at a set distance from the beacon, 1 m for the feasibility experiments and 70 cm for the algorithm evaluation. The higher-gain antenna and closer position for the algorithm evaluation were chosen to ensure feasible voltage and power conditions for our implementation of the power-cycling tag; further design optimization could allow operation at lower voltage and power than demonstrated here. Helpers sit on tripods in the same horizontal plane, about 110 cm above the floor, within a region of 2 × 2 m in front of the power beacon antenna (Fig.~\ref{fig:testbed}). The tripods themselves alter the multi-path. We keep the target and beacon fixed and move only the helpers between scenarios, which limits, but does not remove, that effect. All experiments are conducted in an indoor lab/office environment.

The target's converter output is loaded with 30.1 k$\Omega$, the value that maximized the harvested power $V^2/R$ in a load sweep with only the beacon and target present at 80 cm. We measure the output voltage with an oscilloscope through a 10 M$\Omega$ probe, which has a negligible effect on the load.

\begin{table*}[t]
\centering
\begin{minipage}[t]{0.64\textwidth}
  \centering
  \caption{Testbed Parameters}
  \label{tab:testbed}
  \resizebox{\linewidth}{!}{%
  \begin{tabular}{|c|c|}
    \hline
    \textbf{Item} & \textbf{Value} \\ \hline
    Beacon source / amplifier & USRP N210 (GNU Radio) / Mini-Circuits ZX60-P103LN+ \\
    Frequency / power at antenna port & 917 MHz CW / 22 dBm \\
    Beacon antenna & TE S9025PR patch, 5.5 dBic / Pasternack PE51024 8.5 dBic \\
    Tag antenna & Low Power Labs 915 MHz microstrip dipole, 2.15 dBi \\
    RF switch & SKY13418-485LF (SP8T) \\
    Rectifier & 2-stage Dickson, SMS7621-006LF \\
    Load & 30.1 k$\Omega$ \\
    MCU & ESP32-S3; ATtiny20 \\ \hline
  \end{tabular}}
\end{minipage}\hfill
\begin{minipage}[t]{0.33\textwidth}
  \centering
  \caption{Reflecting States}
  \label{tab:reflections}
  \begin{tabular}{|c|c|c|c|}
    \hline
    \textbf{State} & \textbf{Termination} & $\boldsymbol{\phi_k}$ & \textbf{Gap} \\ \hline
    1 & 6.2 nH & $-122^\circ$ & $36^\circ$ \\
    2 & 2.2 nH & $-86^\circ$  & $33^\circ$ \\
    3 & 12 pF  & $-53^\circ$  & $81^\circ$ \\
    4 & 1.8 pF & $28^\circ$   & $42^\circ$ \\
    5 & 1 pF   & $70^\circ$   & $62^\circ$ \\
    6 & open   & $132^\circ$  & $44^\circ$ \\
    7 & 18 nH  & $176^\circ$  & $62^\circ$ \\ \hline
  \end{tabular}
\end{minipage}
\end{table*}

\section{Methodology}

\subsection{Constructive Interference Model}
\label{sec:model}
Let $\mathcal{R}$ be the set of helpers currently in a reflecting state. We take the beacon's line-of-sight (LOS) field at the target as the phase reference, $E_0$. Helper $i$ re-radiates part of the beacon's wave toward the target along the path beacon $\rightarrow$ helper $i$ $\rightarrow$ target. Relative to the LOS path, that path has amplitude $\alpha_i$, which includes $|\Gamma|\approx 1$ and the two-hop path loss and is typically much less than one. Before the helper's own reflection phase is applied, it has phase $\theta_i$, set by the path-length difference. We keep only single-order reflections and treat helpers in the absorbing state as absent. Because the power beacon only generates a CW signal, multi-path only impacts the phase and amplitude of the signal; thus we absorb static environmental multi-path into the constants $\theta_i $ and $E_0$. The field at the target is

\begin{figure}[t]
    \centering
    \includegraphics[width=0.7\linewidth]{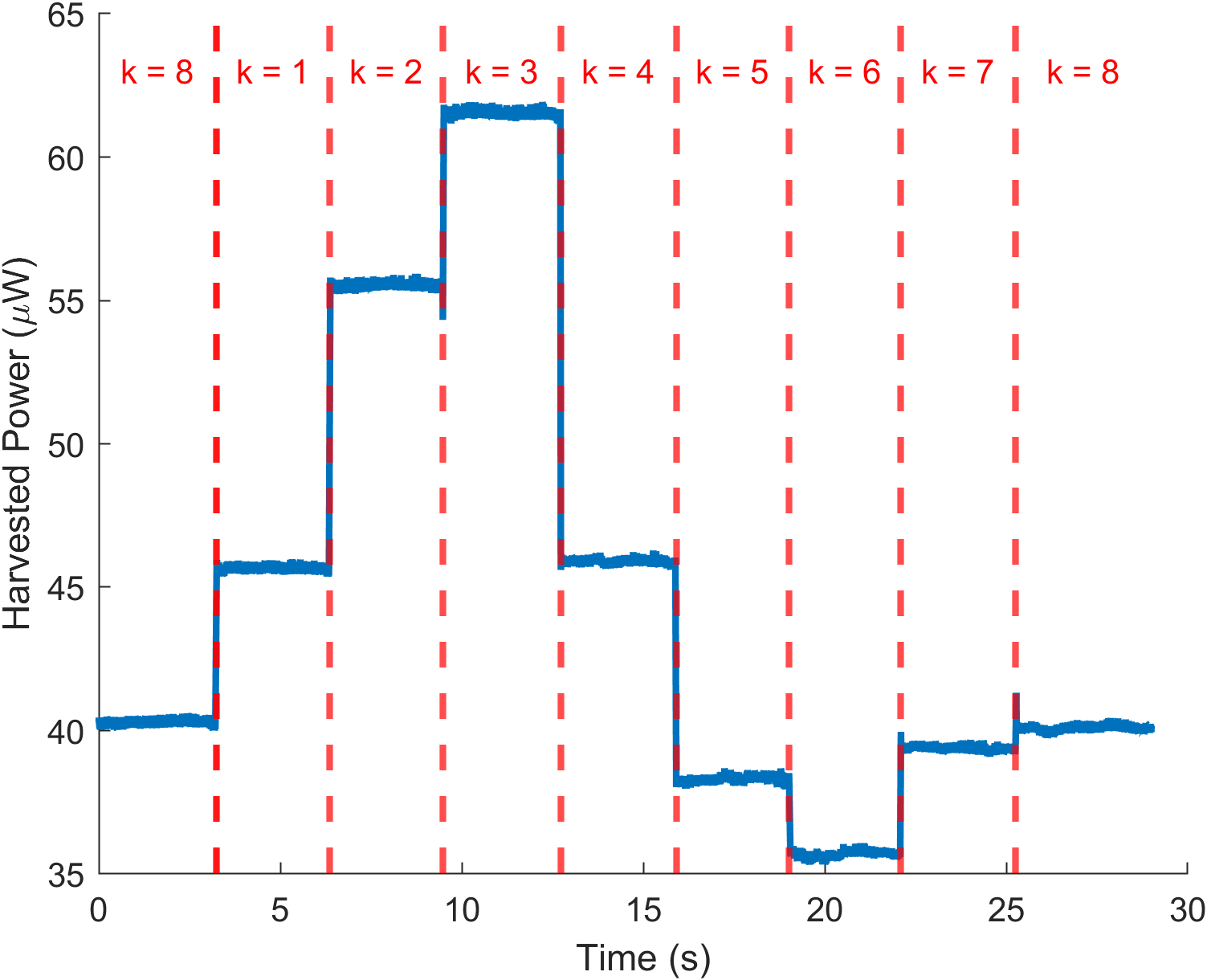}\vspace{-10pt}
    \caption{Harvested power output as the state of a single helper is swept, showing the single-peak response. The helper starts and ends in the absorbing state (k = 8), and sweeps through the reflecting states (k = 1,...,7).}\vspace{-10pt}
    \label{fig:ring order}
\end{figure}

\begin{equation}
    E_T = E_0\Big(1+\sum_{i\in\mathcal{R}}\alpha_i e^{j\psi_i}\Big),\qquad \psi_i=\theta_i+\phi_{k_i},
\end{equation}

where $\phi_{k_i}$ is the reflection phase of the state $k_i$ chosen by helper $i$. The harvested power is a monotonically increasing function of the received power,

\begin{equation}
    P=g\big(|E_T|^2\big).
\end{equation}

Isolating helper $i$ gives

\begin{align}
    &\frac{|E_T|^2}{|E_0|^2}=|A_i|^2+\alpha_i^2+2\alpha_i|A_i|\cos(\psi_i-\angle A_i),\label{eq:4}\\ &A_i=1+\sum_{l\ne i}\alpha_l e^{j\psi_l}.
\end{align}

Three consequences follow, and they underpin the rest of the paper.
\begin{enumerate}
    \item \textbf{Single-peak response.} For fixed $A_i$, power is a sinusoid in $\psi_i$ with one maximum. With discrete states, the measured response is a sampled sinusoid over the ring of states. Fig. \ref{fig:ring order} shows a measurement of this behavior.
    \item \textbf{Nearly independent tuning.} Helper $i$ wants $\psi_i=\angle A_i$, that is, to cancel its own path offset $\theta_i$. That offset depends only on the geometry of beacon, helper $i$, and target. The other helpers enter only through $\angle A_i$, which satisfies $|\angle A_i| \le \arcsin\!\big(\sum_{j\ne i}\alpha_j\big)$ provided $\sum_{j\ne i}\alpha_j<1$. This bound is small when the reflected paths are weak compared to LOS, so the other helpers shift helper $i$'s best phase only slightly. With continuous phase control the global optimum is $\psi_i=0$ for every helper, so the helpers can be tuned in any order without sharing information.
    \item \textbf{A tuned helper always helps the target.} By~\eqref{eq:4}, a reflecting helper increases the target's power relative to the absorbing state whenever its phase error $|\psi_i - \angle A_i|$ is below $90^\circ$, and decreases it when the error is well above $90^\circ$. Because the largest gap between adjacent states in Table~\ref{tab:reflections} is $81^\circ$, the best available state is always within about $41^\circ$ of the ideal phase. A helper set to its best state therefore always increases $P$, retaining at least $\cos 41^\circ \approx 76\%$ of the gain achievable with continuous phase control, whereas a poorly chosen state can reduce $P$ below the absorbing baseline.
\end{enumerate}

\subsection{Sensing the Target's Harvested Power}
The target operates intermittently. It charges a storage capacitor, powers its MCU when a turn-on threshold is reached, transmits a short packet (about 20.5 ms) by OOK-toggling its switch, and returns to charging. The interval $T_p$ between packets is therefore inversely related to the harvested power. A helper can observe $T_p$ without any explicit communication: while in the absorbing state, the target's OOK modulation produces a change in the helper's converter output, which its MCU samples with the ADC. A sample is flagged as a packet when it deviates from a running average by more than a threshold.

\begin{figure}
    \centering
    \includegraphics[width=\linewidth]{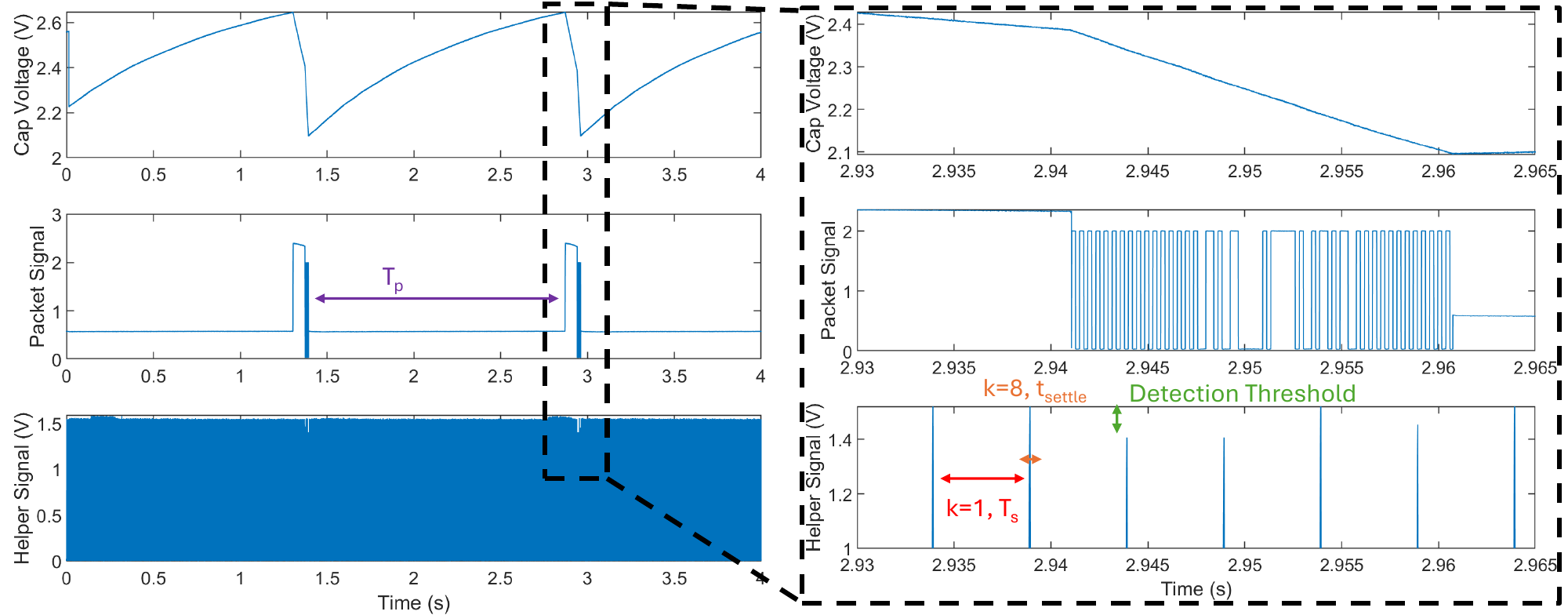}\vspace{-10pt}
    \caption{Timing diagram displaying the capacitor charge-up (top), the burst packet transmission (middle), and the time-multiplexed helper measurements (bottom). A close up image is shown on the right with a label for times when the helper is in the reflecting state (red, k = 1,...,7), and when the helper is in the absorbing state  (orange, k = 8), as well as the difference between the absorption during the packet transmission and when the packet is not being transmitted (green).}\vspace{-10pt}
    \label{fig:Timing Diagram}
\end{figure}

\begin{figure*}[ht]
    \centering
    \subfloat[Geometry 1]{%
        \includegraphics[width=0.32\linewidth]{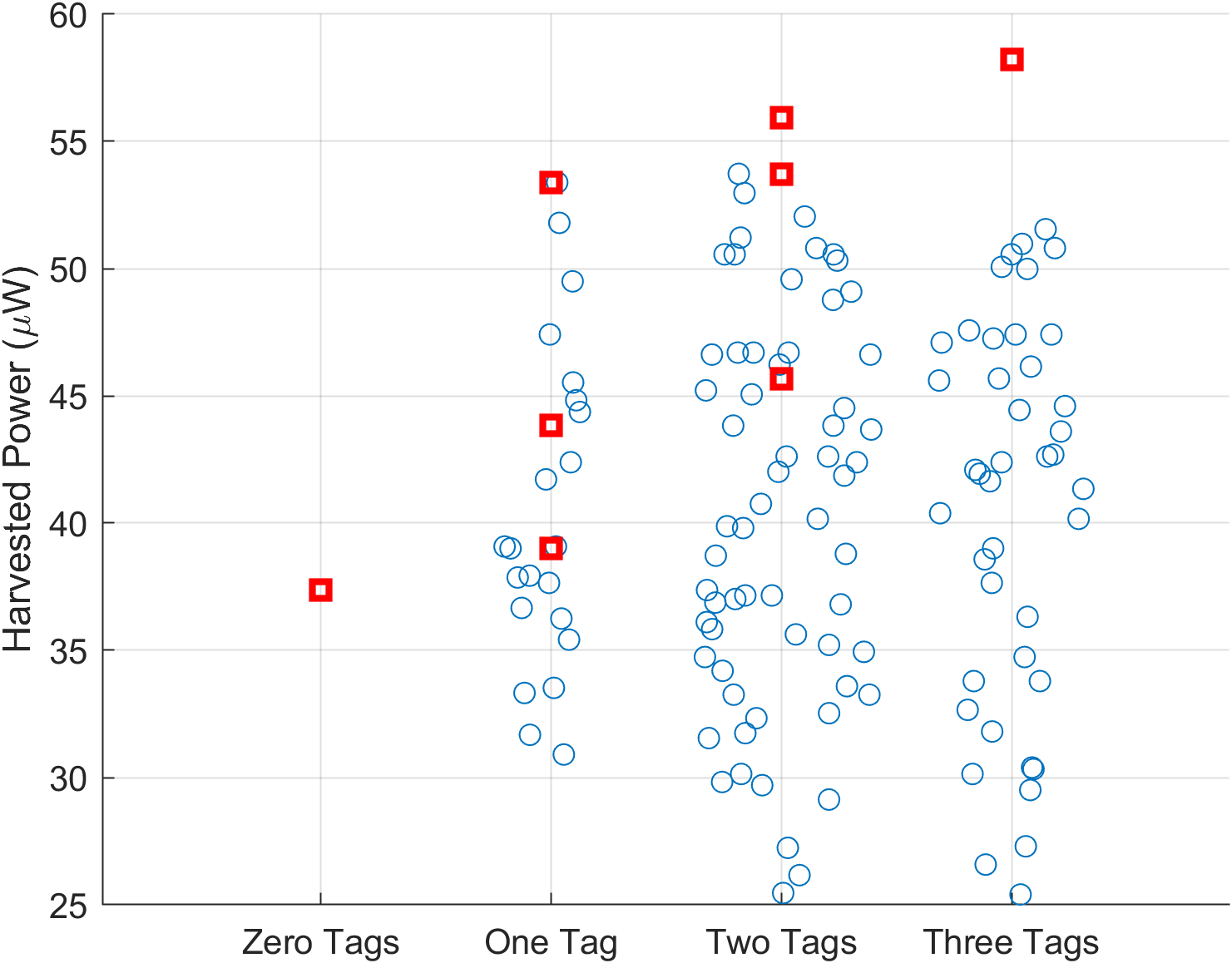}}
    \hfill
    \subfloat[Geometry 2]{%
        \includegraphics[width=0.32\linewidth]{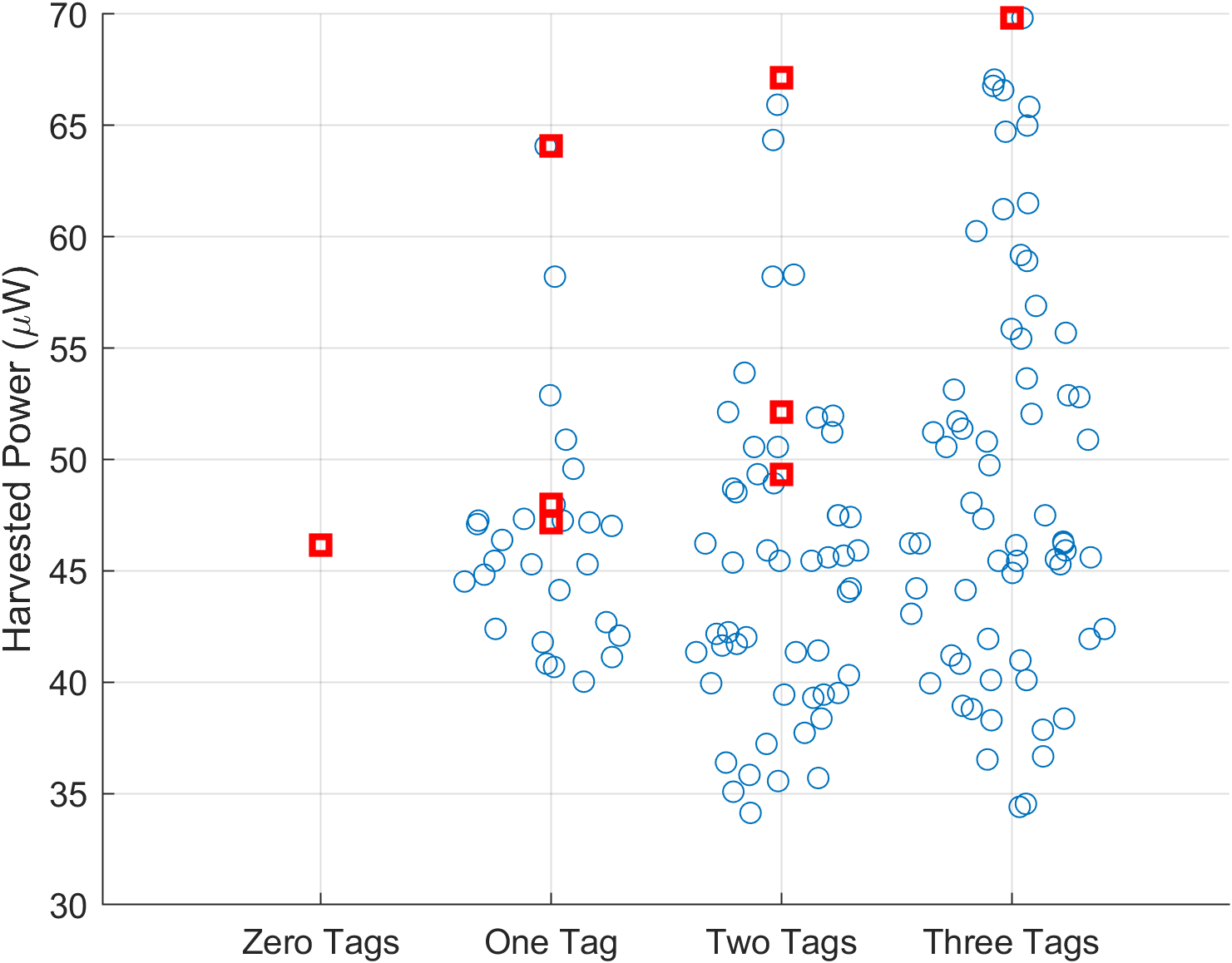}
    }
    \hfill
    \subfloat[Geometry 3]{%
        \includegraphics[width=0.32\linewidth]{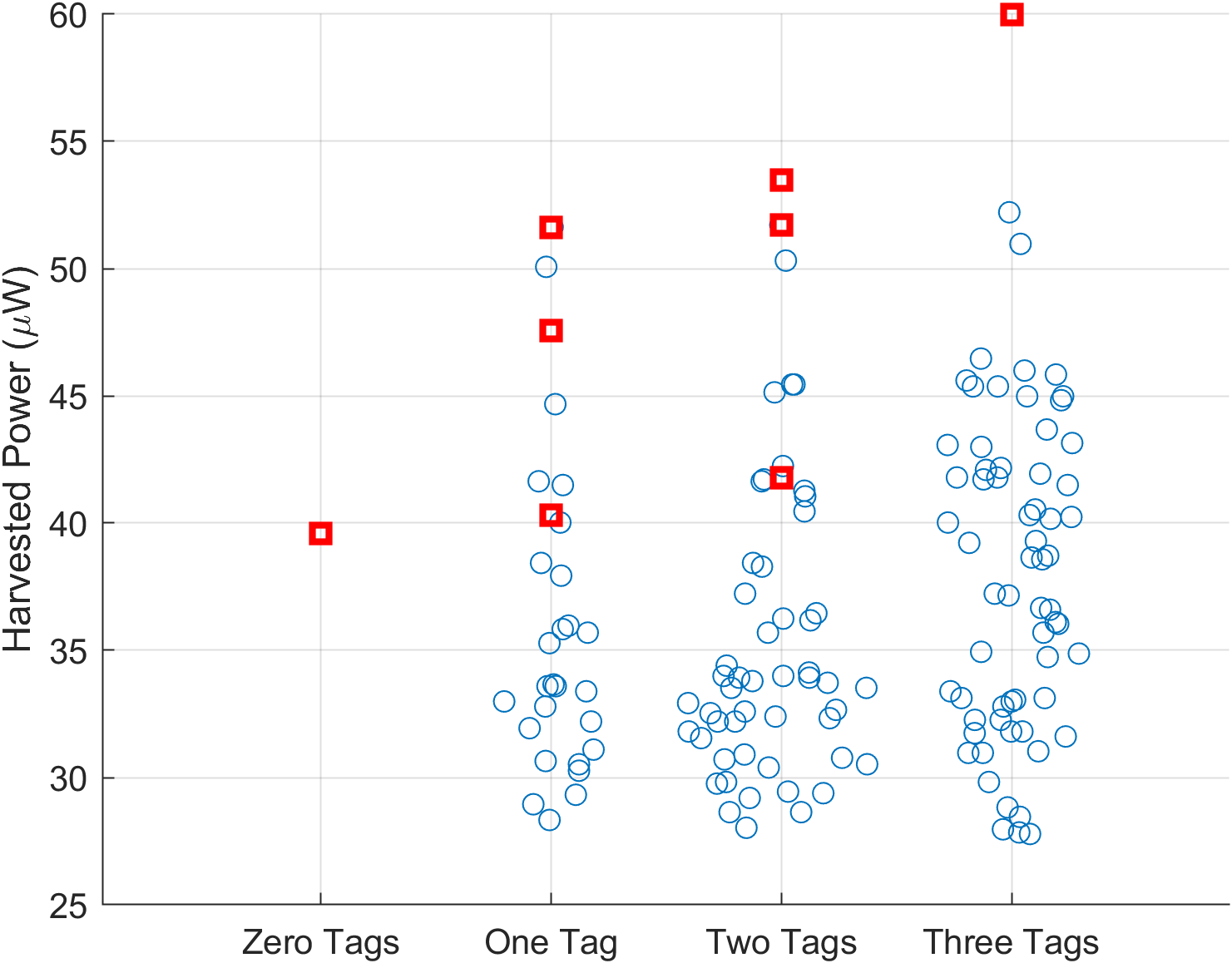}
    }
    \caption{Harvested power improvements from adding differing numbers of tags. Blue circles show the power gained from random impedance values at given number of tags, and red squares show the optimal impedance values. Each plot shows a different physical geometry, with the helper tags in different locations.}\vspace{-10pt}
    \label{fig:feasibility}
\end{figure*}

A tag's antenna can feed only one termination at a time, so a helper cannot listen while it reflects. We therefore time-multiplex. The helper holds its reflecting state and periodically switches to the absorbing state for one sample, waiting long enough for the converter output to settle, which is the time $t_\text{settle}$. Then it returns to reflecting. To avoid missing packets, the sampling period $T_s$ (See $T_p$, $T_s$, and $t_\text{settle}$, labeled in Fig.~\ref{fig:Timing Diagram}) must be shorter than the packet duration. The helper's estimate of $T_p$ is the number of samples $n$ counted between consecutive packets, so its resolution is $T_s$. A smaller $n$ means a higher packet rate and more harvested power. Each state is scored by averaging $n$ over $w$ consecutive intervals, and the first interval after a state change is discarded because it straddles the change. The fraction of time spent in the absorbing state is about $t_{\text{settle}}/T_s$, which reduces the boost the helper can provide. However, the reduction in power boost will still maintain the same relative boost with respect to the given reflecting state. As long as $t_{\text{settle}}/T_s$ is small, the impact will also be marginal.

\subsection{Local Impedance Selection}
\label{sec:algorithm}
Each helper runs the same algorithm (Algorithm \ref{alg:ring-search}), using only its own measurements. States are ordered around a ring by reflection phase (Table~\ref{tab:reflections}), and a state's neighbors are the adjacent states in that order. Because the response is single-peak (Fig.~\ref{fig:ring order}), hill climbing on the ring converges on the individual maximum during that tuning step, and because each tag only takes a single step at a time, the tags will globally converge to a maximum (Section~\ref{sec:algorithm experiment} measures how closely the real-world testbed approaches this). In the noise- and interference- free case, a single tag needs at most three state changes from any starting state.

A helper starts in a random reflecting state, which also supports re-tuning after a tag moves. Its attempts are separated by a random backoff $\tau\sim U(0,W)$, similar to carrier sense multiple access (CSMA). At each attempt it measures the current state ($n_0$) and one neighbor ($n_1$), and it keeps the neighbor only if $n_1<n_0-\delta$. Otherwise it reverts and tries the opposite direction next time. After both directions fail it is at a maximum, so it enlarges $W$ (doubling, up to $W_{\max}$) and checks less often. The helper never stops entirely, so it can adapt to a slow change. There is no communication among helpers. Two helpers whose attempts overlap may misattribute a rate change, but the error is corrected in later attempts, and enlarging $W$ near the optimum makes overlaps rarer.

\begin{algorithm}[]
\small
\caption{Ring Search with Adaptive Backoff}
\label{alg:ring-search}
\begin{algorithmic}[1]
\Require ring-ordered states, window $w$, threshold $\delta$,
         backoff $W_0$, $W_{\max}$
\State $k \gets$ random reflecting state;\ \ $d \gets +1$;\ \ $f \gets 0$;\ \ $W \gets W_0$
\Loop
    \State wait $\tau \sim U(0, W)$
    \State $n_0 \gets \textproc{Measure}(k, w)$
    \State $n_1 \gets \textproc{Measure}(\mathrm{neighbor}(k, d), w)$
    \If{$n_1 < n_0 - \delta$}
        \State $k \gets \mathrm{neighbor}(k, d)$;\ \ $f \gets 0$;\ \ $W \gets W_0$
    \Else
        \State revert to $k$;\ \ $d \gets -d$;\ \ $f \gets f + 1$
        \If{$f \geq 2$}
            \State $W \gets \min(2W, W_{\max})$
        \EndIf
    \EndIf
\EndLoop
\Statex
\Function{Measure}{$k, w$}
    \State hold state $k$
    \State hop to absorbing state every $T_s$
    \State detect packets
    \State discard first interval
    \State \Return median samples between $w$ packets
\EndFunction
\end{algorithmic}
\end{algorithm}

\section{Experimental Results}

\subsection{Feasibility of Helper-Assisted Power Transfer}
\label{sec:feasbility experiment}

\begin{figure*}[ht]
    \centering
    \subfloat[Scenario 1]{%
        \includegraphics[width=0.32\linewidth]{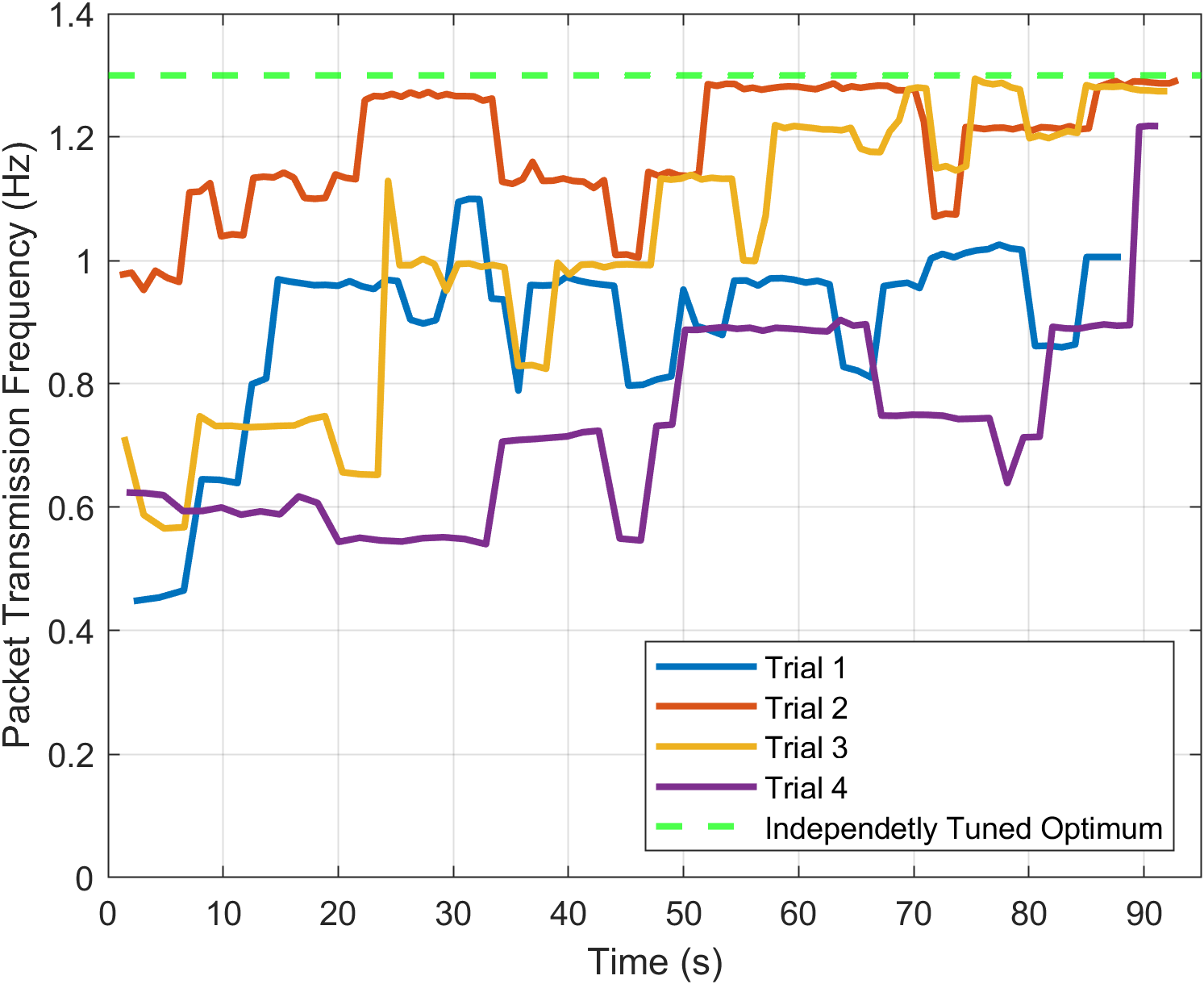}}
    \hfill
    \subfloat[Scenario 2]{%
        \includegraphics[width=0.32\linewidth]{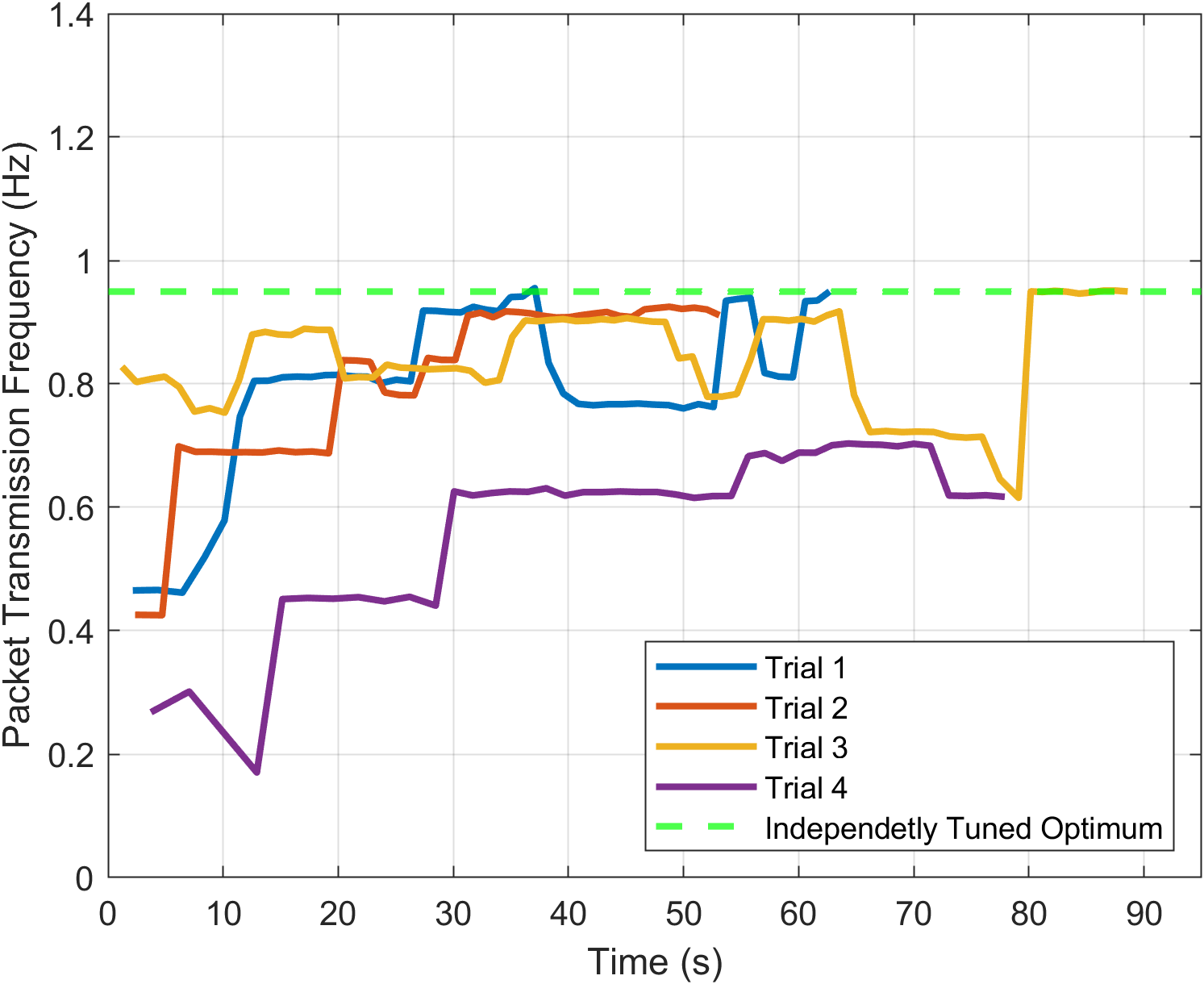}
    }
    \hfill
    \subfloat[Scenario 3]{%
        \includegraphics[width=0.32\linewidth]{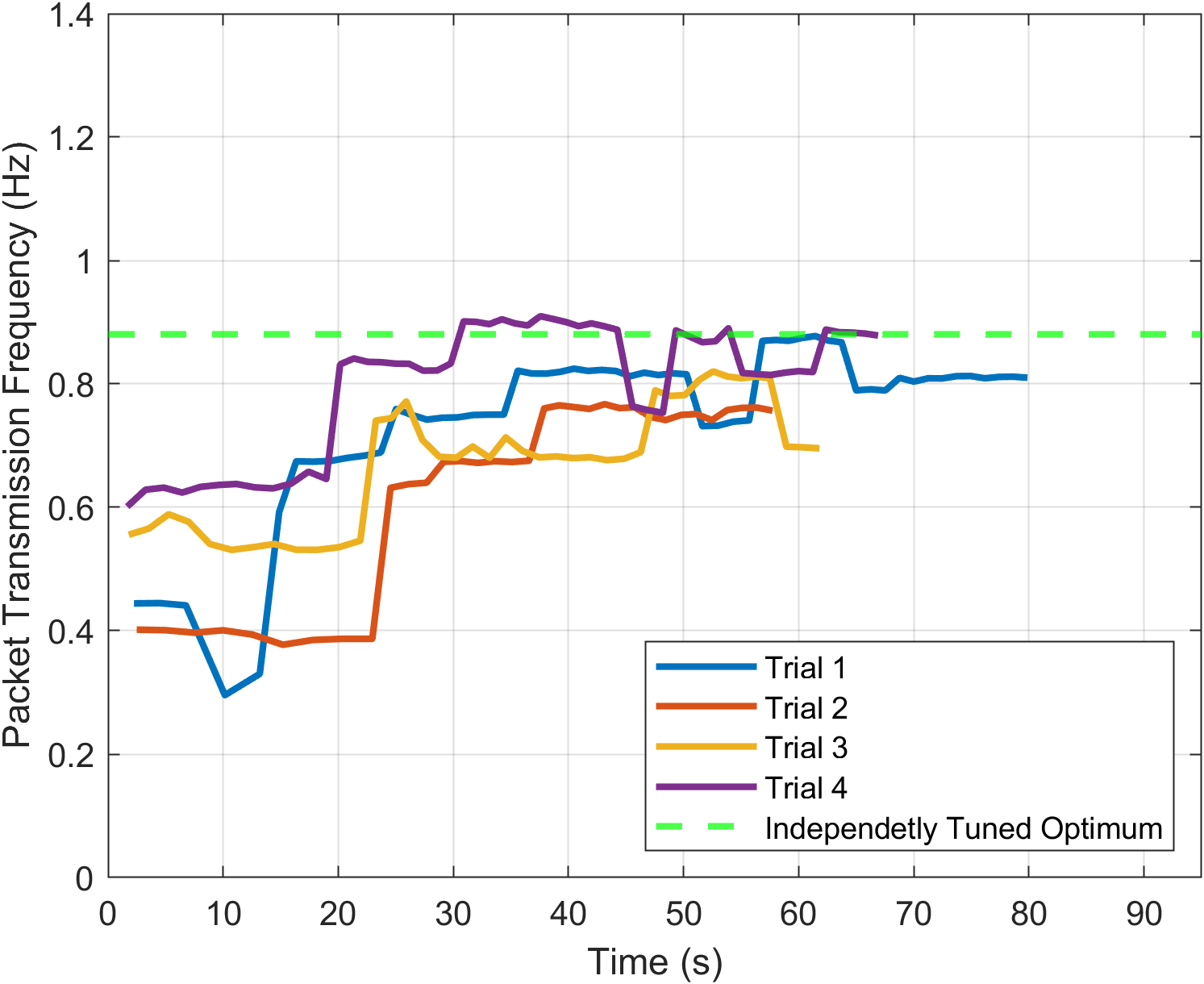}
    }
    \caption{Target tag packet rate over time, showing four trials beginning from randomly set impedance values on each helper tag, and the target rate reached by setting each tag to the independently found optimum impedance.}\vspace{-10pt}
    \label{fig:algorithm}
\end{figure*}

We evaluate the constructive-interference mechanism of Section III-A with the target fixed 1 m from the beacon and up to three helper tags present, using the stage-1 (battery-powered, ESP32-S3 controlled) hardware of Section II. For a given number of active helpers, we measure harvested power across a sample of state combinations, and separately find the best combination for that helper count using the independent per-helper sweep.

Fig. \ref{fig:feasibility} shows harvested power as a function of the number of active helpers for three helper geometries. With no helpers reflecting, the target harvested between 37.37 and 46.15 $\mu$W, with the difference coming from the different multi-paths due to the tripods the helper tags were placed on. For the first geometry, sampling 21 single-helper combinations (one of three helpers reflecting, the other two absorbing) gives a mean of 40.42 $\mu$W, with the optimal helper at the optimal impedance reaching 53.39 $\mu$W (+42.9\% over the baseline); the other two helpers' best states reach 47.41 $\mu$W and 39.01 $\mu$W, showing that each helper sees a different path to the target and thus different maximum power increases. With two helpers active (over 50 sampled combinations across the three possible pairs), the mean is 40.33 $\mu$W and the best pair reaches 55.94 $\mu$W (+49.7\%); the other two pairs' best combinations reach 53.72 $\mu$W and 45.68 $\mu$W. With all three helpers active (42 sampled combinations), the mean is 40.57 $\mu$W and the independently found optimum reaches 58.21 $\mu$W (+55.8\% over baseline). Alternate geometries showed similar boosts to power harvesting, showing that this effect is not dependent on any specific geometry.

Two patterns are visible. First, the independently tuned configuration's power rises with the number of participating helpers (37.37 $\rightarrow$ 53.39 $\rightarrow$ 55.94 $\rightarrow$ 58.21 $\mu$W), consistent with each additional correctly-phased reflection adding constructively. Second, the mean of randomly sampled combinations stays flat at roughly 40 $\mu$W regardless of how many helpers are active, only modestly above the baseline. Random reflection is as likely to interfere destructively as constructively, so simply adding more reflectors without phase control yields little benefit and may even be detrimental; the gain comes specifically from setting each helper's state, not from the presence of additional reflecting tags.

At one and two helpers the plot shows multiple optimal points (three at one helper, three at two helpers), one per distinct subset of helpers. This reflects that different helpers, and different pairs, sit on different paths to the target and so reach different peak gains; we vary this geometry across trials without separately isolating its effect.

\subsection{Local Impedance Selection}
\label{sec:algorithm experiment}

We evaluate Algorithm 1 on the stage-2 (power-cycling) target at a fixed distance of 70 cm, using an 8.5 dBic beacon antenna and the parameters in Table~\ref{tab:parameters}. We test three helper placements (Scenarios 1 to 3, distinct from the geometries of Section \ref{sec:feasbility experiment}) with four independent trials each, and report every trial. In each trial the three helpers start from random states and run Algorithm 1 with no external intervention. We record the target’s storage-capacitor voltage with an oscilloscope, independent of the helpers’ own sensing, and take the packet rate as the inverse of the interval between successive charge cycles. As a reference for each scenario, we hold each helper at its individually tuned state and measure the target’s rate. This tuned optimum is the dashed line in Fig. 7.

In every trial the rate ends higher than it starts. Trials start at a median of 50\% of the tuned optimum (range 28\% to 87\%) and end at a median of 95\% (range 65\% to 100\%). That is a median improvement of 1.9$\times$ in packet rate, with 8 of the 12 trials ending at 90\% or more of the optimum. A few traces briefly exceed the reference (up to 103\%), which we attribute to small multi-path differences between the reference measurement and the trial.

The traces show the structure of the hill climb. The rate rises in steps as helpers accept moves, with short dips lasting a few packets where a helper has moved to a neighboring state to test it and then reverted. The trace shows only the combined effect of the three helpers, so we do not attribute individual steps to individual helpers. How quickly the rate approaches the optimum is set mainly by the parameters in Table~\ref{tab:parameters}. Each state-change attempt requires two packet measurements, at roughly 7 s per state change at the rates observed here, after a random wait of up to $W_0 = 15$ s, so each helper completes only a few state change attempts in a trial. Shorter waits and windows would speed this up at the cost of more overlap between helpers’ measurements (Section \ref{sec:algorithm}).

\begin{table}[ht]
\caption{Algorithm Parameters}
\begin{center}
\begin{tabular}{|c|c|}
\hline
\textbf{Parameter}&\textbf{Value}\\
\hline
Window size $w$	& 3\\
Acceptance threshold $\delta$ &	2\\
Initial backoff $W_0$	& 15 s\\
Maximum backoff $W_\text{max}$	& 180 s\\
Sampling period $T_s$	& 5 ms\\
Settling time $t_\text{settle}$	&8 $\mu$s\\
Target distance	&70 cm\\
\hline
\end{tabular}
\label{tab:parameters}
\end{center}
\end{table}

\section{Conclusion \& Future Work}
We present a method for improving RF wireless power transfer in battery-less backscatter sensor networks by having nearby tags act as helpers, using discrete impedance control to constructively combine their reflected signal with the beacon's line-of-sight path at a target tag. On a hardware testbed, phase-optimized helpers increased the target's harvested power by up to 55.8\% over an all-absorbing baseline, while randomly set helper phases gave no reliable benefit, confirming that the gain comes from phase control rather than merely adding reflectors. We further presented a local, autonomous algorithm that lets each helper tune its own impedance using only measurements of the target's packet rate, without channel estimation or coordination between helpers, and showed that it raises the target’s packet rate to a median of 95\% of that of the independently tuned configuration.

These results indicate that passive, independent sensor nodes can meaningfully extend and improve each other's power-transfer capabilities without significant added infrastructure, additional beacons, or centralized control, a property well suited to dense, swarm-style battery-less sensor deployments. Future work includes characterization of the loss introduced by the discrete reflection states used here, relative to finer- or coarser- grained phase control, and extending the algorithm to tags that reposition slowly over time.
\bibliographystyle{IEEEtran}
\bibliography{references}

\end{document}